\documentclass[conference]{./IEEEtran}
\usepackage{cite}

\usepackage{soul}

\ifCLASSINFOpdf
  \usepackage[pdftex]{graphicx}
\else
  \usepackage[dvips]{graphicx}
\fi
\usepackage{amsmath}
\usepackage{algorithm}
\usepackage{algorithmic}
\usepackage{amsfonts,amssymb,amsthm}
\usepackage{upgreek}
\usepackage[tight]{subfigure}
\usepackage{multirow}
\usepackage{tabularx}

\newcommand\blfootnote[1]{%
  \begingroup
  \renewcommand\thefootnote{}\footnote{#1}%
  \addtocounter{footnote}{-1}%
  \endgroup
}

\begin{document}

%

\title{MAC-Oriented Programmable Terahertz PHY via Graphene-based Yagi-Uda Antennas}

%
%
%



%
\author{\IEEEauthorblockN{Seyed Ehsan Hosseininejad\IEEEauthorrefmark{1}\IEEEauthorrefmark{5},
Sergi Abadal\IEEEauthorrefmark{2},
Mohammad Neshat\IEEEauthorrefmark{1}, 
Reza Faraji-Dana\IEEEauthorrefmark{1}, 
Max C. Lemme\IEEEauthorrefmark{3},\\
Christoph Suessmeier\IEEEauthorrefmark{4}, 
Peter Haring Bol\'{i}var\IEEEauthorrefmark{4}, 
Eduard Alarc\'{o}n\IEEEauthorrefmark{2}, 
and Albert Cabellos-Aparicio\IEEEauthorrefmark{2}}
\IEEEauthorblockA{\IEEEauthorrefmark{1}School of Electrical and Computer Engineering\\
University of Tehran, Tehran, Iran}%
\IEEEauthorblockA{\IEEEauthorrefmark{2}NaNoNetworking Center in Catalonia (N3Cat), \\
Universitat Polit\`{e}cnica de Catalunya, Barcelona, Spain}%
\IEEEauthorblockA{\IEEEauthorrefmark{3}Faculty of Electrical Engineering and Information Technology, \\
RWTH Aachen University, Aachen, Germany}
\IEEEauthorblockA{\IEEEauthorrefmark{4}Institute for High Frequency Electronic and Quantum Electronics, \\
University of Siegen, Siegen, Germany}
}

\maketitle

\begin{abstract}
Graphene is enabling a plethora of applications in a wide range of fields due to its unique electrical, mechanical, and optical properties. In the realm of wireless communications, graphene shows great promise for the implementation of miniaturized and tunable antennas in the terahertz band. These unique advantages open the door to new reconfigurable antenna structures which, in turn, enable novel communication protocols at different levels of the stack. This paper explores both aspects by, first, presenting a terahertz Yagi-Uda-like antenna concept that achieves reconfiguration both in frequency and beam direction simultaneously. Then, a programmable antenna controller design is proposed to expose the reconfigurability to the PHY and MAC layers, and several examples of its applicability are given. The performance and cost of the proposed scheme is evaluated through full-wave simulations and comparative analysis, demonstrating reconfigurability at nanosecond granularity with overheads below 0.02 mm\textsuperscript{2} and 0.2 mW. 


\end{abstract}


%
\IEEEpeerreviewmaketitle

\section{Introduction}
\label{sec:intro}
%
%
%
%

\blfootnote{\IEEEauthorrefmark{5}Email: sehosseininejad@ut.ac.ir}Terahertz (THz) wireless communication has attracted increasing interest due to the demand for extremely high data rates envisaged for future networks. The THz band could provide a broad bandwidth with low area and power footprints, therefore enabling a multitude of applications the such as cellular networks beyond 5G, terabit local or personal area networks, or secure military communications \cite{Akyildiz2014}.


An efficient antenna is crucial to satisfy the stringent performance and cost requirements set by multi-Gbps links in the THz band. Graphene is an excellent candidate for the implementation of THz antennas owing to its ability to support the propagation of surface-plasmon polaritons (SPPs) in this particular frequency range \cite{xiao2016graphene}. Since SPPs are \emph{slow waves}, graphene antennas show a considerable miniaturization potential \cite{Hosseininejad2016}. SPPs are also tunable, thereby providing graphene plasmonic devices with unique reconfiguration capabilities \cite{Hosseininejad2017}. Graphene-based antennas can not only be frequency-agile when graphene is used as the radiating element, but also introduce adaptive matching or beam reconfigurability \cite{Correas2017}. Thus far, however, reconfigurability has been only explored in one direction at a time. 


The unique properties of graphene antennas open the door to radically new applications and communication protocols. On the one hand, miniaturization turns graphene antennas to a perfect fit for area-constrained applications such as wireless networks among nanosensors \cite{Akyildiz2014}, within chips \cite{Abadal2017}, or within programmable metamaterials \cite{AbadalACCESS}. On the other hand, reconfigurability could be used to address challenges faced in the design of Physical (PHY) layer and Medium Access Control (MAC) protocols for THz networks, where directionality leads to the deafness problem \cite{Xia2015} or molecular absorption leads to highly distance-dependent transmission windows \cite{Han2014a}. However, although several THz PHY/MAC works have assumed the use of graphene antennas \cite{Han2014a, Xia2015}, very few have exploited their unique characteristics or analyzed their potential impact at higher layers of design \cite{Lin2014, Han2017}.

%
\begin{figure*}[!t] 
\centering
\subfigure[Yagi-Uda plasmonic graphene antenna (Sec. II).\label{fig:Yagi}]{\includegraphics[width=0.9\columnwidth]{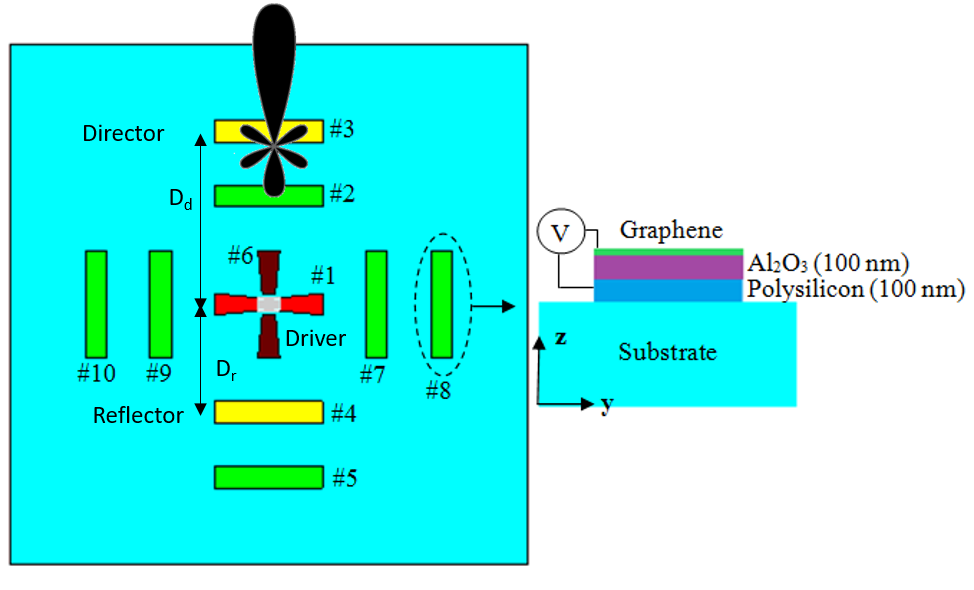}}
\subfigure[Programmable antenna controller (Sec. III).\label{fig:scheme}]{\includegraphics[width=0.9\columnwidth]{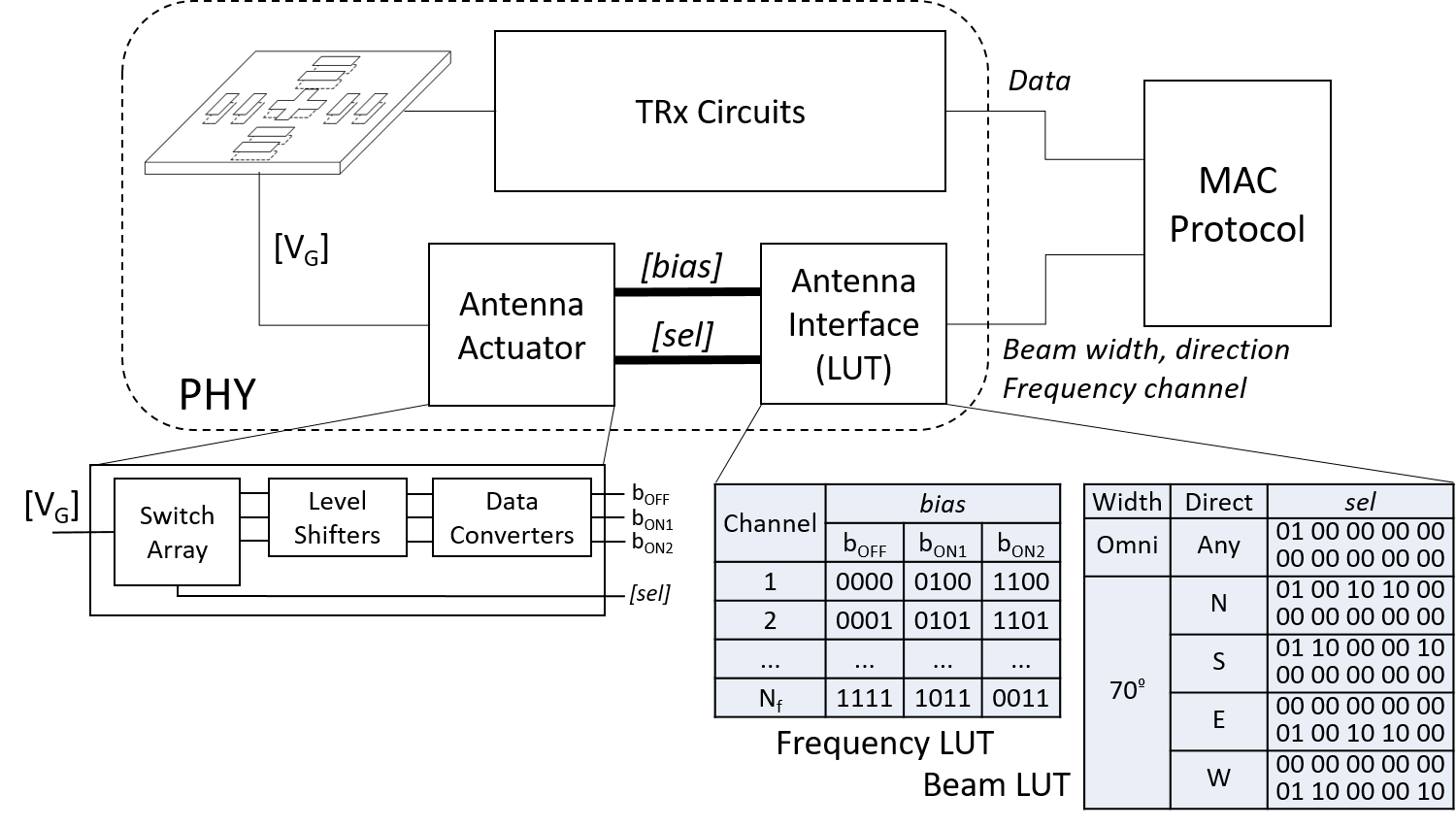}}
\vspace{-0.1cm}
\caption{Overview of the contribution of this work: programmable PHY at THz frequencies with graphene-based Yagi-Uda antennas.}
\vspace{-0.5cm}
\end{figure*}

In this paper, we bring both ends together by (1) exploring a graphene antenna design that delivers frequency and beam reconfiguration simultaneously, and (2) present a smart antenna controller that exposes the reconfigurability capabilities to the PHY and MAC layers. The frequency--beam agility of the proposed antenna ensures its applicability in a wide range of scenarios, whereas the controller provides flexibility and ease of use with high reconfiguration speed. 

Our proposed THz antenna is based on the classical Yagi-Uda antenna, which consists of a linear dipole array driven by a single feed element. This concept was first explored in \cite{han2010terahertz}, where the Yagi-Uda structure is scaled to achieve the same mechanism of directional emission in the THz band and designed with a high input resistance to improve impedance matching with typical THz sources. Others \cite{wu2016graphene, Xu2014MIMO} have explored the role of graphene sheets in Yagi-Uda antennas to change the radiation pattern while attaining miniaturization. Our scheme employs graphene both in the driven and parasitic elements, joining frequency adaptivity with beam agility in a compact structure. In this work, we evaluate an instance of such an antenna and demonstrate its beam reconfiguration in two different frequency channels.  

On the controller side, we consider a scheme composed by an array of actuators driven by a digital interface and outline its design methodology. The design is conceptually similar to that of \cite{Rodrigo2016} at microwave frequencies, but easier to integrate and adapt to support any bias-controlled THz graphene antenna. We discuss the voltage range and resolution requirements at the controller, which mainly depend on the biasing structure, and investigate their impact on the performance of the antenna. We also quantify the overheads involved in the design, seeking to demonstrate that the simplicity of the scheme allows for faster reconfiguration than with conventional designs. 


The rest of this paper is organized as follows. 
Section \ref{sec:yagi-uda} describes the Yagi-Uda plasmonic graphene antenna, whereas Section \ref{sec:programmable} outlines the design principles of the antenna controller. Section \ref{sec:evaluation} evaluates the antenna performance and explores the tradeoffs and overheads of the antenna controller. Then, Section \ref{sec:MACdesign} discusses the potential uses of our scheme with relevant link-level examples. Finally, Section \ref{sec:conclusion} concludes the paper.

\section{A Graphene-based Reconfigurable Yagi-Uda Antenna}
\label{sec:yagi-uda}
Yagi-Uda antennas are very famous for RF and microwave systems because of their relatively high directivity and simplicity. The basic unit of this kind of antenna is an array consisting of at least three dipole elements. The central one is driven element and determines the frequency of operation, whereas the rest are parasitic elements that are used to improve directivity. Note that the higher the number of parasitic elements, the higher the directivity. 

Here, we propose a THz antenna structure based on graphene using idea of Yagi-Uda antenna to achieve reconfigurability both in frequency and radiation beam. The main idea is to use graphene both in the driver and the parasitic elements and to dynamically tune their chemical potential to provide the required frequency and beam. Basically, the frequency depends on the chemical potential at the driver and the beam is shaped by the chemical potential at the parasitic elements. 
 
\subsection{Antenna Geometry} 
\label{sec:geometry}
Fig. \ref{fig:Yagi} depicts a schematic representation of the proposed THz antenna with the geometrical parameters which possess the reconfigurable ability in both the working frequency and radiation pattern. It consists of two orthogonal driven dipoles and four parasitic strips made of graphene. This four-beam reconfigurable antenna consists of four three-element Yagi-Uda antennas such that all the dipoles are composed of few-layer graphene. To change the chemical potential of graphene, a DC voltage is applied between a 100-nm thick polysilicon layer and the graphene strip using a 100-nm thick Al$_{2}$O$_{3}$ spacer. The substrate is considered high resistivity Gallium Arsenide (GaAs). The values of chemical potentials of the central and surrounding dipoles determine which one will affect the antenna performance. 


\subsection{Design Decisions} 
\label{Yagi}
To design the proposed antenna, the main step is designing the basic three-element Yagi-Uda antenna. The elements of this unit including driver, director, and reflector should be arranged to form a directional array antenna in THz frequencies. In the conventional Yagi-Uda antennas, the parasitic elements are considered somewhat smaller or longer in length than the feed element, while in the proposed antenna, the graphene dipole elements are of the same length. To achieve the end-fire beam formation in this structure, the graphene dipole elements are \emph{electrically} shortened or lengthened by applying different chemical potentials. 

According to feeding approach of antenna, first resonance (low-impedance) or second resonance (high-impedance) of driven dipole should be chosen to achieve impedance matching \cite{Tamagnone2012}. Here, we consider second one in order to reduce the return loss when connecting the plasmonic antenna to a THz source, which will typically have high impedance. Consequently, a full-wavelength dipole, which operates in the high-impedance resonance is used as a driver. Regarding classical antenna theory, positioning a parasitic element close to driver produce an end-fire radiation. If the parasitic element, here graphene plasmonic element, has greater chemical potential compared to the driven graphene dipole, it acts like a reflecting element. If the parasitic element is positioned at large enough spacing related to driver, it can act as a directing element. 

Finally, we replicate the basic unit in order to the cover the full azimuth plane using four beams. The direction of main beam is determined by the chemical potentials applied to the driven and parasitic elements. The elements will be in conductive state (tuned in) in the presence of a non-zero chemical potential, or in dielectric state (tuned out) otherwise.

\subsection{Modeling and Evaluation Methodology} 
To evaluate the proposed structure, it is necessary to model the conductivity of graphene at terahertz frequencies. The Kubo formalism is generally employed for single-layer graphene \cite{Correas2017}, leading to
\begin{equation}
\sigma\left(\omega\right)=\frac{2e^{2}}{\pi\hbar}\frac{k_{B}T}{\hbar}\ln\left[2\cosh\left[\frac{E_{F}}{2k_{B}T}\right]\right]\frac{i}{\omega+i\tau^{-1}},\label{eq:sigma_graphene}
\end{equation}
where $e$, $\hbar$, and $k_{B}$ are constants corresponding to the charge of an electron, the reduced Planck constant, and the Boltzmann constant, respectively. Variables $T$, $E_{F}$, and $\tau$ correspond to the temperature, the chemical potential, and the relaxation time of the graphene layer. For $N<6$ graphene layers, the optical conductivity can be approximated as $\sigma_{N} = N \sigma$ \cite{Casiraghi2007}. 

The frequency--beam reconfigurability of the proposed antenna is achieved by tuning the conductivity via changes in the chemical potential $E_{F}$, either through electrostatic biasing or chemical doping \cite{Huang2012ARRAY, Gomez2015}. The relaxation time depends on the quality of the graphene sheets and generally takes values of hundreds of femtoseconds. Once the conductivity is known, graphene can be modeled as an infinitesimally thin sheet with surface impedance $Z = 1/\sigma_{N}$.

\section{Programmable PHY at THz Frequencies}
\label{sec:programmable} 
The tunability of plasmons in graphene opens the door to elegant solutions for programmable PHY in the THz band. With the Yagi-Uda antennas discussed above, one can determine the frequency of resonance, beam width, and beam direction by simply changing the bias applied to the different graphene sheets. Here, we propose a simple controller that exposes such reconfigurability potential to the upper layers.

Figure \ref{fig:scheme} outlines the proposed scheme in an example that exploits the Yagi-Uda antenna reconfigurability at the MAC level. The tunable antenna is connected to both the transceiver, which we assume to be tunable as well \cite{Jornet2014TRANSCEIVER}, and a controller. The controller translates directives coming from the MAC protocol (in this particular example) to a set of voltages that determine the antenna state (frequency and beam). An antenna interface and an antenna actuator, which are described next, are required to this end.

\subsection{Antenna Interface}
The interface of the antenna controller is composed by two Look-Up Tables (LUTs) that translate the antenna state requirements into digital signals that drive the actuator. The reconfigurability process has two steps. 

In the first step, the frequency channel is translated into codes representing various levels of \emph{bias}. Our proposed antenna only requires three levels: $b_{ON1}$ for the driven element, $b_{ON2}$ for the reflectors and directors, and $b_{OFF}$ for the tuned off elements. However, the scheme can be generalized to any antenna structure with any number of levels. The number of bits of the \emph{bias} signal depends on the range of voltages and the resolution required to implement a given number of channels to a given graphene antenna.

In the second step, the beam width and direction are translated into a code \emph{sel} that identifies the \emph{bias} level to apply to each antenna element. In the example of Fig. \ref{fig:scheme}, $b_{OFF} \leftarrow 00$, $b_{ON1} \leftarrow 01$ and $b_{ON2} \leftarrow 10$ and the antenna elements are addressed as numbered in Fig. \ref{fig:Yagi}. 

\begin{table}[!b] 
\vspace{-0.5cm}
\caption{Geometry of the evaluated antenna at 2.3 THz.}
\label{tab:param}
\vspace{-0.2cm}
\footnotesize
\centering
\begin{tabular}{|c|c|c|c|c|c|} 
\hline
Parameter         & Description & Value \\ \hline
$L$ ($\upmu$m)     	& Length of dipole elements & $25$  \\ \hline
$D_d$ ($\upmu$m) 		& Distance from director to central dipole	& $40$ \\ \hline
$D_r$ ($\upmu$m)		& Distance from reflector to central dipole  & $25$	\\ \hline
\end{tabular}
\end{table}

\subsection{Antenna Actuator}
The actuator uses the signals coming from the actuator to deliver the appropriate voltage $v_{g}(i) \in [V_{G}]$ to each antenna element. The actual voltage value depends on factors such as the biasing structure, the number of graphene layers, or the required chemical potential \cite{Huang2012ARRAY, Gomez2015}. The biasing configuration assumed in this work is similar to that used in \cite{Huang2012ARRAY}, which gates graphene through a thin Al$_{2}$O$_{3}$ layer. In such configuration and assuming single-layer graphene, the chemical potential variation $\Delta E_{F}$ relates to the change of voltage $\Delta v_{g}$ as 
\begin{equation}
\label{eq:SLG}
\Delta v_{g}^{SLG} = \frac{e E_{F}^2 t}{\pi \hbar^2 v_{F}^2 \epsilon_{0} \epsilon_{r}},
\end{equation}
where $e$ is the elementary charge, $\hbar$ is the reduced Planck constant, $v_{F} \approx 10^{6}$ is the Fermi velocity, $\epsilon_{0}$ is the vacuum permittivity, whereas $\epsilon_{d}$ and $t$ are the permittivity and thickness of the material below graphene \cite{Yu2009Chemical}. The relation $v_{g}$--$E_{F}$ becomes linear for bilayer graphene and remains unclear for trilayer graphene and above. In any case, the scheme here presented is applicable to any graphene structure capable of being reconfigured through electrostatic biasing.


\section{Performance and Cost Evaluation}
\label{sec:evaluation}
We use CST \cite{CST} to model and evaluate the antenna of Fig. \ref{fig:Yagi}. We illustrate its beamforming capabilities in Section \ref{beam} and the joint frequency--beam reconfigurability in Section \ref{freq--beam}. Through reasonable assumptions, we estimate the potential impact of the antenna controller limitations upon the antenna characteristics in Section \ref{impact}. Overheads are estimated and discussed in Section \ref{overheads}.

\begin{table*}[!t] 
\footnotesize
\caption{Chemical potentials (in eV) of the evaluated antenna at 2.3 THz.}
\label{tab:chemical}
\vspace{-0.2cm}
\centering
\begin{tabular}{|c|cccccccccc|}
\hline
Pattern & $E_{F1}$ & $E_{F2}$ & $E_{F3}$ & $E_{F4}$ & $E_{F5}$ & $E_{F6}$ & $E_{F7}$ & $E_{F8}$ & $E_{F9}$ & $E_{F10}$ \\ \hline \hline
Omnidirectional & 0.5 & 0 & 0 & 0 & 0 & 0.5 & 0 & 0 & 0 & 0 \\ \hline
+Y direction & 0.5 & 0 & 0.8 & 0.8 & 0 & 0 & 0 & 0 & 0 & 0 \\ \hline 
-Y direction & 0.5 & 0.8 & 0 & 0 & 0.8 & 0 & 0 & 0 & 0 & 0 \\ \hline
+X direction & 0 & 0 & 0 & 0 & 0 & 0.5 & 0 & 0.8 & 0.8 & 0 \\ \hline
-X direction & 0 & 0 & 0 & 0 & 0 & 0.5 & 0.8 & 0 & 0 & 0.8 \\ \hline
\end{tabular}
\vspace{-0.3cm}
\end{table*}

\begin{figure*}[!t] 
\centering
\subfigure[Omni, XY plane\label{fig:single-xy}]{\includegraphics[width=0.48\columnwidth]{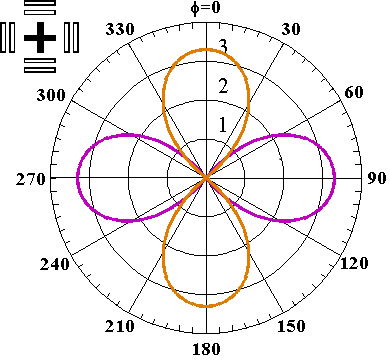} }
\subfigure[Omni, YZ plane\label{single-yz-xz}]{\includegraphics[width=0.48\columnwidth]{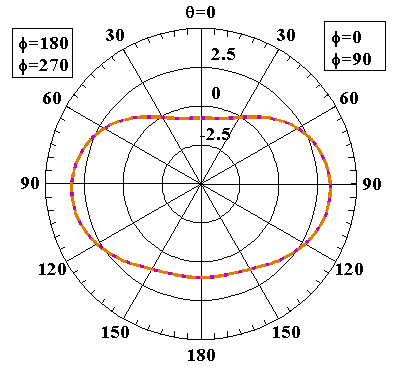} }
\subfigure[Direct, XY plane\label{fig:yagi-xy}]{\includegraphics[width=0.48\columnwidth]{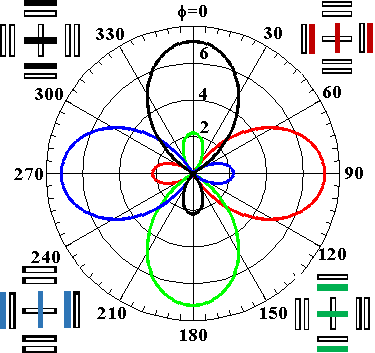} }
\subfigure[Direct, YZ plane\label{yagi-yz-xz}]{\includegraphics[width=0.48\columnwidth]{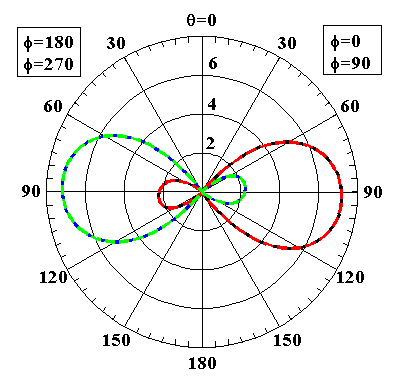} }
\vspace{-0.1cm}
\caption{Directivity patterns of the antenna in omnidirectional (a-b) and directional modes (c-d).}
\label{fig:single}
\vspace{-0.3cm}
\end{figure*}

\subsection{Beam Reconfigurability}
\label{beam}
According to the rules explained in Section \ref{sec:yagi-uda}, we estimate a set of design points and then use CST to obtain locally optimal values for the chemical potential in a given frequency. Table \ref{tab:param} shows the values for the proposed antenna. To have the Yagi-Uda array for +Y-direction beam, the chemical potential of driven dipole (element 1) is set to 0.5 eV, whereas the chemical potentials of the elements 3 and 4 (see Fig. \ref{fig:Yagi}) are set to 0.8 eV such that they behave as a director and a reflector considering their distance from the central dipole. Notice that the other elements are set to 0 eV. Due to XY symmetry, it is straightforward to calculate the chemical potentials for the rest of radiation beams. Table \ref{tab:chemical} summarizes their values.

Let us now compare the omnidirectional and directional modes. On the one hand, Figures \ref{fig:single-xy} and \ref{single-yz-xz} illustrate the directivity patterns at 2.3 THz, which corresponds to the high-impedance resonance point of the structure, in its omnidirectional mode for the XY and YZ (XZ) planes. The maximum gain is 1.6 dB, whereas the radiation efficiency is 40\%. On the other hand, Figures \ref{fig:yagi-xy} and \ref{yagi-yz-xz} illustrate the patterns of the four radiation beams, also at 2.3 THz and for the XY and YZ (XZ) planes. The use of the reflector and director improves the gain to 5.2 dB with a -3dB beamwidth of 69\textsuperscript{o} front-to-back ratio of 4.9 dB, thus demonstrating a remarkable beam reconfigurability.



\subsection{Joint Frequency-Beam Reconfigurability}
\label{freq--beam}
The frequency tunability of graphene-based antennas has been demonstrated in numerous works \cite{Correas2017}. Here, we instead focus on illustrating the joint frequency--beam reconfigurability with an example. To this end, let us consider a Yagi-Uda structure incorporating a third parasitic element placed 75 $\upmu$m away from the central dipole. This structure is capable to work at either 1.5 THz or 2.3 THz. For the second working point, we keep the chemical potential configuration considered in Table \ref{tab:chemical}. For the new working point at 1.5 THz, we consider a chemical potential of 0.2 eV at the driven element and 0.5 eV at the parasitic elements. The dipoles located 40 $\upmu$m away from the driven element are reused and will be the reflectors in this new frequency; the newly included elements 75 $\upmu$m away from the driver will be the directors. 

\begin{figure}[!t] 
\centering
\includegraphics[width=0.8\columnwidth]{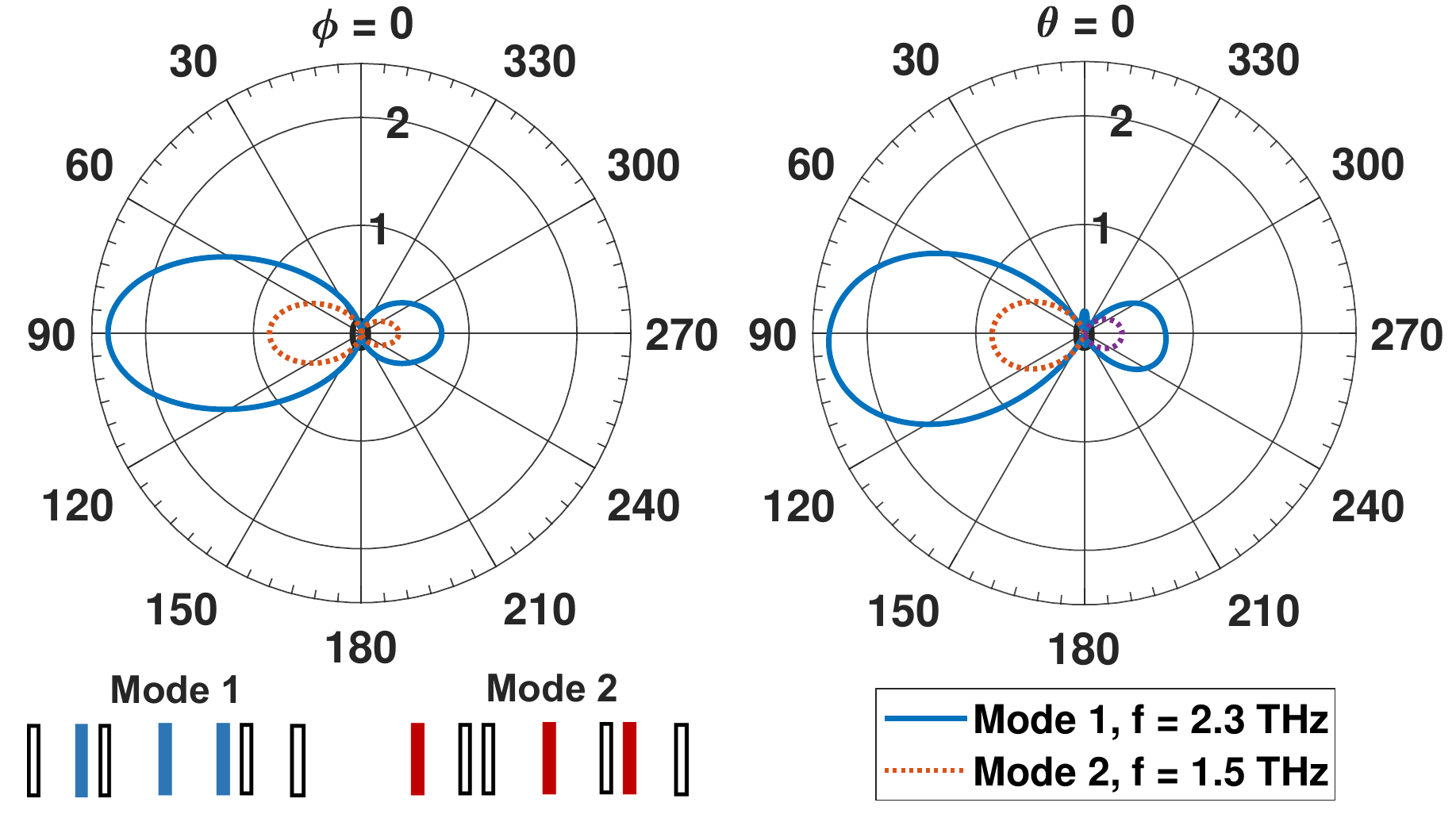}
\vspace{-0.25cm}
\caption{Directivity patterns in the XY plane (left) and YZ plane (right) at 1.5 and 2.3 THz.}
\label{fig:joint}
\vspace{-0.2cm}
\end{figure}

Figure \ref{fig:joint} shows a comparison of the directive pattern at the two working frequencies. At 1.5 THz, the gain is more modest due to the lower chemical potential applied to the elements, but still improves from -5.13 dB to -0.73 dB with a beamwidth of 73\textsuperscript{o} and a front-to-back ratio of 4.17 dB, therefore demonstrating the versatility of the proposed scheme.


\subsection{Performance Impact of the Antenna Controller}
\label{impact}
The antenna controller may have a limited range (max $\Delta v_{g}$) and resolution ($\Delta v_{g}$/step) due to device limitations in area or energy-constrained applications. The former will constrain the number of available frequency channels, whereas the latter may introduce undesired effects when tuning out parasitic elements that are not required in a given directional mode. Here, we perform a preliminary evaluation of such effects.

For simplicity, let us consider the scheme of Fig. \ref{fig:Yagi} with single-layer graphene in the driven and parasitic elements and a source with impedance $Z_{S} = 1 k\Omega$. We use Equation \eqref{eq:SLG} with $t = 100$ nm and $\epsilon_{r} = 9.3$ to quantify the voltage variation required to produce a given change in chemical potential, to then obtain the shift of the antenna resonance produced by such variation. Using the return loss at -10 dB to calculate the bandwidth, we finally obtain the number of non-overlapping channels achievable with a given voltage range. 

\begin{figure}[!t] 
\centering
\includegraphics[width=0.9\columnwidth]{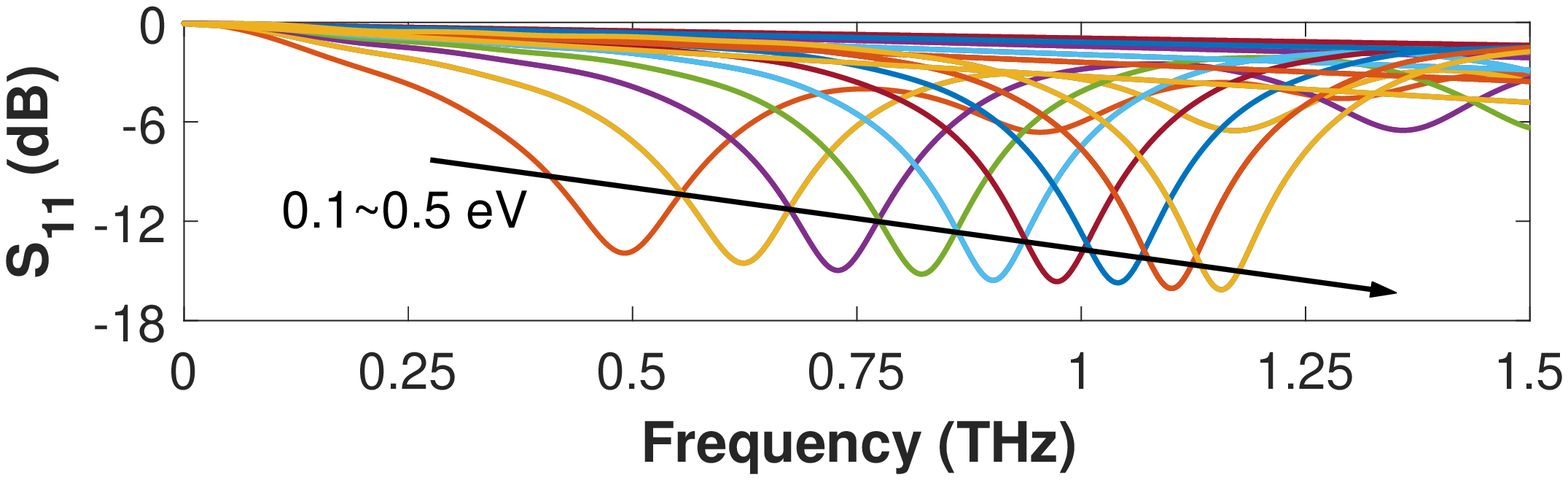} 
\vspace{-0.4cm}
\includegraphics[width=0.42\columnwidth]{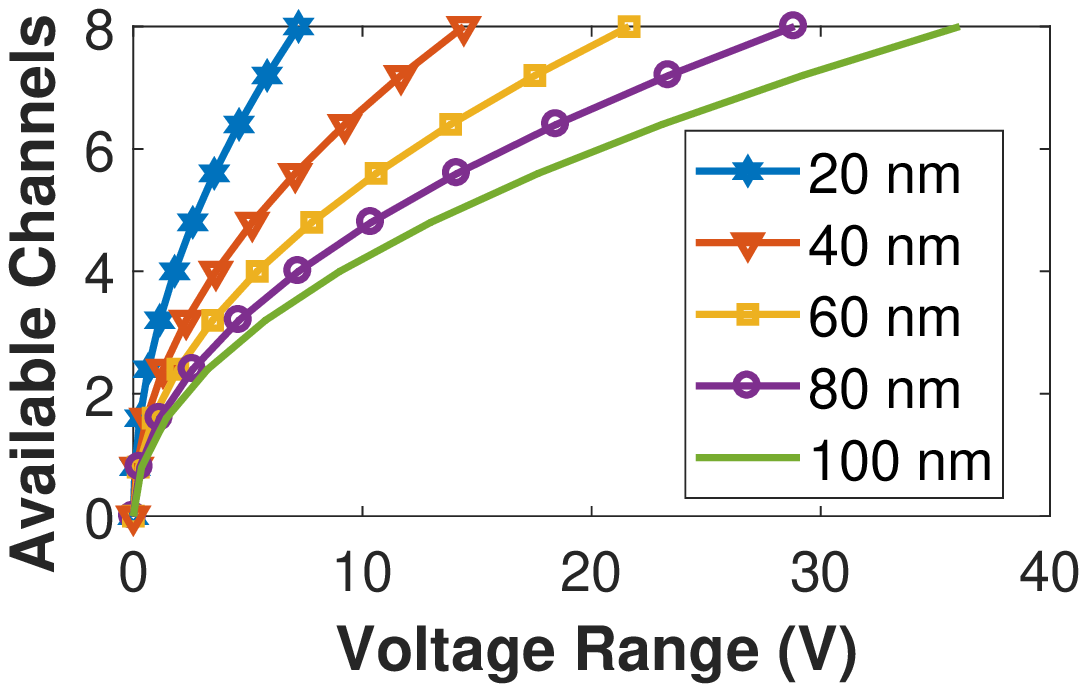}
\includegraphics[width=0.42\columnwidth]{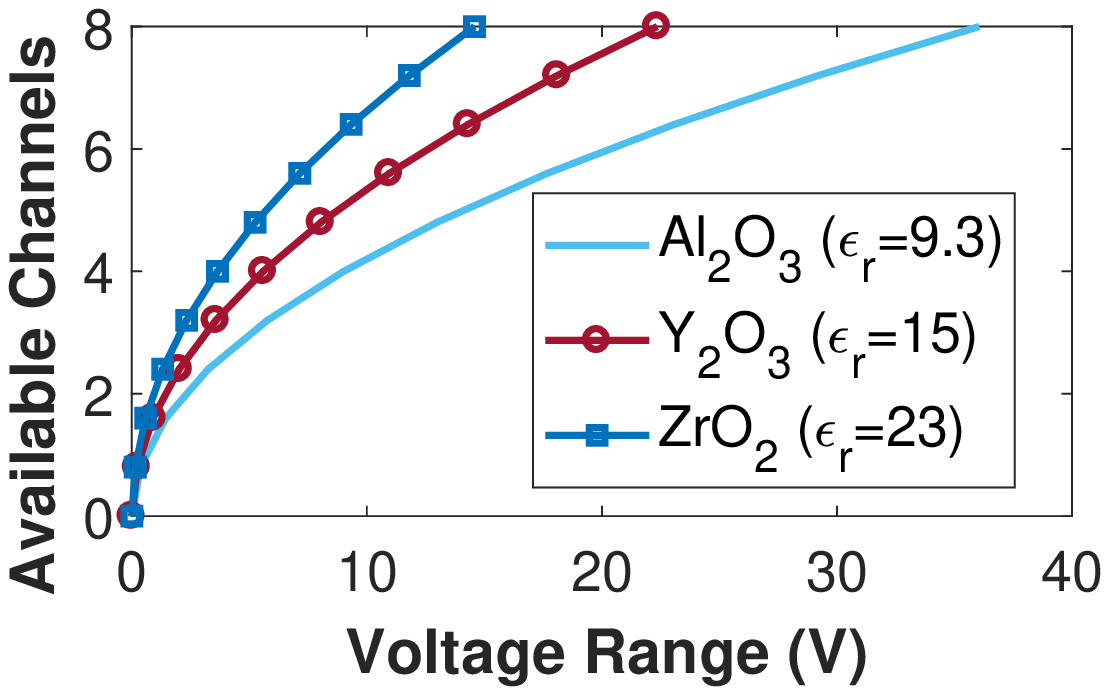}
\caption{Scaling of resonance point (top) and of the available channels as function of the voltage range for different spacer thicknesses and dielectric constants (bottom).}
\label{fig:a}
\vspace{-0.3cm}
\end{figure}

Figure \ref{fig:a} shows the number of available channels as a function of the voltage range of our proposed antenna. With a voltage range of $\sim$35 V, we can achieve a chemical potential variation of up to 0.5 eV, which leads to a resonant frequency of $\sim$1.2 THz with a bandwidth of 140 GHz as shown in the top chart of Fig. \ref{fig:a}. This means that up to 8 disjoint frequency channels can be potentially considered. Since the chemical potential scales as $O(v_{g}^2)$, it becomes increasingly difficult to open new channels as we stretch the voltage range. To enlarge the practicable spectrum with the same voltage budget, one can reduce the thickness of the spacer or use dielectric materials with higher index (bottom charts of Fig. \ref{fig:a}) \cite{Robertson2004}. This, however, increases the complexity and cost of the gating structure.

To assess the effects of a controller with limited voltage resolution, let us consider a non-ideal gating scheme that is not capable of fully tuning off the elements of the antenna \cite{Huang2012ARRAY}. This implies that tuned off elements will still have a residual conductivity. We model such effect by defining their conductivity as a fraction of the conductivity of the driven element and evaluate the structure of Fig. \ref{fig:Yagi} at 2.3 THz.

\begin{figure}[!t] 
\centering
\includegraphics[width=0.8\columnwidth]{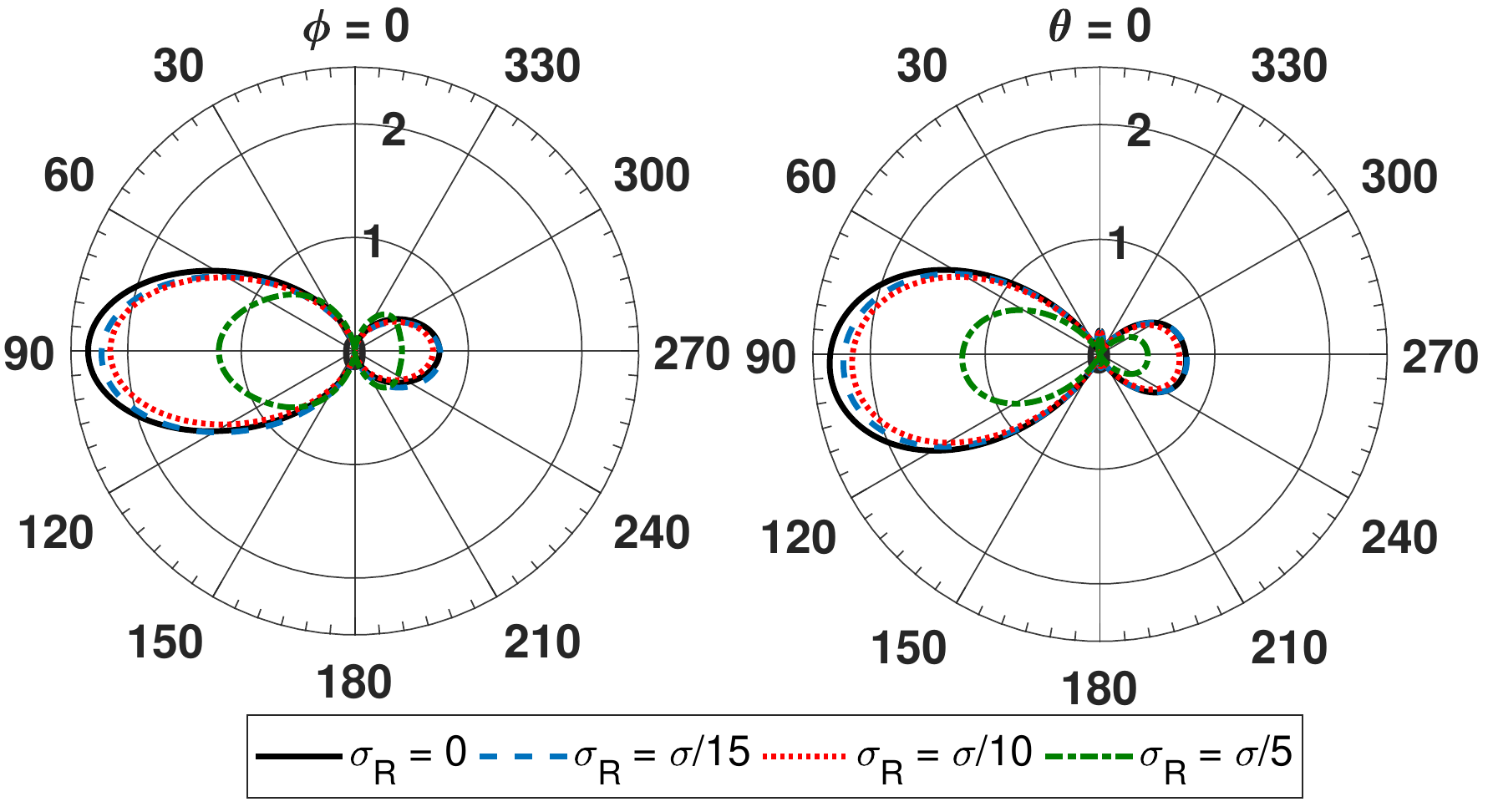} 
\vspace{-0.25cm}
\caption{Directivity pattern in the XY plane (left) and YZ plane (right) for different levels of residual conductivity $\sigma_R$.}
\label{fig:b}
\vspace{-0.3cm}
\end{figure}

Figure \ref{fig:b} illustrates the impact of an increasing residual conductivity to the directivity patterns of the antenna. Fortunately, the reconfigurability potential is maintained intact with a front-to-back ratio of around 5 dB. However, the antenna performance is reduced as the gain drops from 2.35 dB to 0.93 dB and the beamwidth increases from 70\textsuperscript{o} to 92\textsuperscript{o} when the residual conductivity is one fifth than that of the driven element. For this particular antenna, a residual conductivity of one fifteenth can already be considered safe.

\subsection{Antenna Controller Overhead}
\label{overheads}
In resource-limited applications such as wireless networks for nanosensors \cite{Akyildiz2014} or within reconfigurable metasurfaces \cite{AbadalACCESS}, any area or energy overhead can be crucial. Here,  the main sources of area and power consumption are the LUTs and the data converters. To estimate the former, we model the LUTs as a single cache of 2KB with a line size of 32 bits and use CACTI \cite{CACTI} to derive their cost. At 32nm, this memory would be 90$\times$40 $\upmu$m\textsuperscript{2} and consume less than 0.02 nJ to read a configuration, with a leakage power of less than 10 $\upmu$W. Despite of the small overhead, this antenna interface would be capable of retaining as many as 512 antenna states (our examples have a few dozens) with fairly large resolution. To model the data converters and level shifters, we refer to the literature. In the former case, Kim \emph{et al.} design a 90-nm data converter with 6-bit resolution capable of working at 3.1 GS/s \cite{Kim2016DAC}. By conservatively scaling down to 4 bits and 1 GS/s, enough to support 16 frequency channels and reconfigurability at nanosecond scale, the area and power become 0.01 mm\textsuperscript{2} and 0.15 mW at 32nm. Compared with the area and power of an RF transceiver or the complexity of phase shifters or delay lines required for MIMO, these overheads are negligible.

A similar analysis can be performed in terms of delay. All the components of the antenna controller, including the data converter, can work at the nanosecond scale with reasonable area and power overheads. The change of graphene conductivity required to reconfigure the antenna also occurs in a sub-ns scale \cite{Huang2012ARRAY}, turning this solution into a much faster alternative to the conventional beamsteering techniques. 


\section{Discussion: Potential Use Cases}
\label{sec:MACdesign}

\noindent \textbf{Link-level support for DAMC.}
Distance-Aware Multi-Carrier (DAMC) modulation is a recent proposal for terahertz band communication at distances from meters to kilometers \cite{Han2014a}. DAMC selects the transmission windows depending on the distance between nodes because, at longer distances, undesired attenuation peaks appear in the channel response due to molecular absorption. This requires a method to estimate the channel, which is not defined. 

In our case, exposing the antenna reconfigurability to the link layer allows to keep track of the distance between mobile transmitters and receivers with low overhead. Nodes can agree on a common \emph{safe} channel and regularly switch to it, estimate the transmission distance, and then switch to a new data channel with a new directivity level if required.

\vspace{0.2cm}
\noindent \textbf{Multichannel MAC Protocols in the THz band.}
To date, a few MAC protocols for THz networks have been proposed. The performance of conventional MAC protocols when upscaled has been discussed in \cite{AbadalMAC}, but those works fail to adapt to the particularities of the THz band. In the macroscale scenario, the use of highly directional antennas is required to combat the large propagation loss, which leads to the deafness problem at the MAC layer. To address this, Xia \emph{el al.} propose a protocols where receivers broadcast availability by sweeping the surrounding space, but do not provide means to perform this function \cite{Xia2015}. Others have proposed to derive handshaking to a lower band where omnidirectional modes can be used safely, but this option requires nodes to be equipped with multiple radios \cite{Yao2016}. An alternative uses omnidirectional antennas in the service discovery phase, to then switch to highly directional antennas for data transmission \cite{Han2017}.

To the best of the authors knowledge, the only works assuming unique antenna reconfigurability are \cite{Lin2014, Han2017}. Their approach considers that the beam can be reconfigured at the pulse level and schedules transmissions from the access point (AP) accordingly. Such assumption implies beam switching at sub-ns granularity, but the authors do detail how to do it. 

Our proposed scheme is not only capable of achieving the beam switching capability required by \cite{Lin2014}, but could also be applied to works requiring beam scanning \cite{Xia2015} or a control channel at lower frequencies with an omnidirectional antenna \cite{Yao2016, Han2017}. Moreover, protocols could be extended to use multiple data channels at disjoint frequencies. The pseudocode for a generalized example of such asynchronous multichannel MAC with centralized coordination is shown in Algorithm \ref{alg1} from the transmitter's perspective. We assume that the AP keeps track of the network topology and active transmissions, and attaches the channel and direction to use to the CTS.

\begin{algorithm}[!t]
\caption{Simplified multichannel MAC protocol using the antenna controller (controller instructions in bold).}
\label{alg1}
\begin{algorithmic}
\small
\STATE \textbf{set(omni, any\_dir, control\_ch)}
\STATE send(RTS, addrRX, control\_ch)
\STATE \{data\_ch, direction\} $\leftarrow$ receive(CTS, addrRX, control\_ch)
\STATE \textbf{set(70\textsuperscript{o}, direction, data\_ch)}
\STATE send(data, addrRX, data\_ch)
\end{algorithmic}
\end{algorithm}

%
%
%


\section{Conclusion}
\label{sec:conclusion}
We have presented a design for programmable PHY in the THz band, detailing and evaluating its two key elements: a frequency--beam reconfigurable Yagi-Uda antenna based on graphene and a suitable antenna controller. The proposed antenna supports five beam configurations with 4-dB directivity enhancement at two distinct frequencies, but the practicable configuration space can be extended by placing more parasitic elements around the driver. The antenna controller exposes the reconfigurability to the upper layers at nanosencond granularity and negligible overheads (0.01 mm\textsuperscript{2} and 0.15 mW at 32nm), opening the door to MAC protocols with unprecedented versatility and speed.



%



\section*{Acknowledgment}
This work has been funded by Iran's National Elites Foundation (INEF), the Spanish Ministry \emph{Econom\'ia y Competitividad} under grant PCIN-2015-012, the European Union under grant H2020-FETOPEN-736876, and the German Research Foundation (DFG) grants HA 3022/9-1 and LE 2440/3-1.

\end{document}